\documentclass[aps,prb,groupedaddress,amsmath]{revtex4}

\usepackage[usenames]{color}

\def\etal{\mbox{et al.}}

\usepackage{amssymb}
\usepackage{amsmath}
\usepackage{graphicx}
\usepackage{amsbsy}

\newcommand{\bq}{\begin{eqnarray}}
\newcommand{\eq}{\end{eqnarray}}
\newcommand{\bqn}{\begin{eqnarray*}}
\newcommand{\eqn}{\end{eqnarray*}}

\newcommand\beq{\begin{equation}}
\newcommand\eeq{\end{equation}}
\newcommand\beqa{\begin{eqnarray}}
\newcommand\eeqa{\end{eqnarray}}
\newcommand{\nn}{\nonumber\\}

\begin{document}

\title{The penetrable square-well model: extensive versus non-extensive phases}

\author{Riccardo Fantoni$^{a}$$^{\ast}$\thanks{$^\ast$Email: rfantoni@ts.infn.it
\vspace{6pt}},
Alexandr Malijevsk\'y$^{b}$$^{\dagger}$\thanks{$^\dagger$ Email: a.malijevsky@imperial.ac.uk
\vspace{6pt}},
Andr\'es Santos$^{c}$$^{\ddagger}$\thanks{$^\ddagger$Email: andres@unex.es
\vspace{6pt}}
and Achille Giacometti$^{d}$$^{\S}$\thanks{$^\S$Corresponding author. Email: achille@unive.it
\vspace{6pt}}
\\
\vspace{6pt}  $^{a}${\em{National Institute for Theoretical Physics (NITheP) and Institute of Theoretical Physics, Stellenbosch 7600, South Africa}};\\
$^{b}${\em{E. H{\'a}la Laboratory of Thermodynamics, Institute of  Chemical Process Fundamentals of the ASCR, And Department of
  Physical Chemistry, Institute of Chemical Technology, Prague, 166 28  Praha 6, Czech Republic}};\\
$^{c}${\em{Departamento de F\'isica, Universidad de Extremadura,  E-06071 Badajoz, Spain}};\\
  $^{d}${\em{Dipartimento di Scienze Molecolari e Nanosistemi,  Universit\`a Ca' Foscari Venezia, S. Marta DD2137, I-30123 Venezia, Italy}}
}

\begin{abstract}
The phase diagram of the penetrable square-well fluid is investigated
through Monte Carlo simulations of various nature. This model was proposed as the simplest possibility
of combining bounded repulsions at short scale and short-range attractions. We prove that the model is
thermodynamically stable for sufficiently low values of the
penetrability parameter, and in this case the system behaves similarly to the
square-well model. For larger penetration, there exists an intermediate region
where the system is metastable, with well defined fluid-fluid and fluid-solid transitions, at finite
size, but eventually becomes unstable in the thermodynamic limit. We
characterize the unstable non-extensive phase appearing at high
penetrability, where the system collapses into an isolated blob of a few
clusters of many ovelapping particles each.

\end{abstract}

\maketitle
\section{Introduction}
\label{sec:introduction}
Unlike simple fluids, complex fluids are typically characterized by a significant reduction in the number of degrees of freedom, in view of the hierarchy of different length and energy scales involved. As a result,
coarse-grained potentials accounting for effective interactions between a pair of the complex fluid units adopt analytical forms that are often quite different from those considered paradigmatic for simple fluids
\cite{Likos01}.

An important example of this class of potentials is given by those  bounded
at small separations, thus indicating the possibility of a partial (or even total) interpenetration.
This possibility, completely unphysical in the framework of simple fluids, becomes on the contrary
very realistic in the context of complex fluids. While the true two-body interactions
always include a hard-core part, accounting for the fact that energies close to contact raise
several orders of magnitude, effective interactions obtained upon averaging microscopical degrees of freedom
may or may not present this feature, depending on the considered particular system.

Interesting examples {with no hard-core part} are given by polymer solutions,
where effective polymer-polymer interactions can be argued to be of the
Gaussian form \cite{Stillinger76,Bolhuis01,McCarty10}, and star polymers and dendrimers where the so-called penetrable sphere (PS) model
is frequently employed \cite{Marquest89,Likos98,Mladek08}.

In spite of their markedly different phase behaviors \cite{Mladek08}, both these effective interactions have the common {attributes of being} bounded at zero separation and {lacking} an attractive part. The latter
feature, however, appears to be particularly limiting in view of the several sources of attractive interactions typical of polymer solution, such as, for instance, depletion forces \cite{Bolhuis01}, that are
typically accounted through simple attractive square-well (SW) tails.

A tentative of combining both the penetrability at small separation and the attraction at slighty larger scale,
led to the introduction of the penetrable square-well (PSW) potential \cite{Santos08,Fantoni09,Fantoni10a,Fantoni10b,Mukamel2009}.
This can be obtained either by starting from the PS model and adding an attractive well, or by starting
from the SW model and reducing the infinite repulsive energy to a finite one. In this way, the model
is characterized by two length scales (the soft core and the width of the well) and by two energy scales,
the height $\epsilon_r$ of the repulsive barrier  and the depth $\epsilon_a$ of the attractive well.

The ratio $\epsilon_a/\epsilon_r$, hereafter {simply} referred to as {``penetrability''}, is a measure of the accessibility of the repulsive barrier and, as we shall see, plays a very important role in the equilibrium
properties of the fluid. When $\epsilon_a/\epsilon_r=0$, the PSW model reduces to the {PS model (if $k_BT/\epsilon_r=\text{finite}$, where $T$ is the temperature) or to the SW model (if $k_BT/\epsilon_a=\text{finite}$). In
the latter case, the model} exhibits a fluid-fluid phase transition for any width of the attractive square well \cite{Vega92,deMiguel97,delRio02,Liu05,Giacometti09}, this transition becoming metastable  against the formation
of the solid for a sufficiently narrow well \cite{Liu05}.  As penetrability $\epsilon_a/\epsilon_r$ increases, different particles tend to interpenetrate more and more because {this becomes} energetically favorable (the
precise degree depending on the $\epsilon_a/\epsilon_r$  ratio). As a result, the total energy {may grow} boundlessly to
negative values and the system can no longer be thermodynamically stable. The next question to be addressed is whether this instability occurs for any infinitesimally small value $\epsilon_a/\epsilon_r>0$ or, conversely,
whether there exists a particular value where the transition from stable to unstable regime occurs.

As early as the late sixties, the concept of a well-behaving thermodynamic limit was translated into a simple rule, known as Ruelle's criterion \cite{Fisher66,Ruelle69}, for the sufficient condition for a system to be stable.
In a previous paper \cite{Santos08}, we have discussed the validity of Ruelle's criterion for the one-dimensional PSW case and found that, indeed, there is a well-defined value of penetrability $\epsilon_a/\epsilon_r$, that
depends upon the range of the attractive tail, below which the system is definitely stable. Within this region, the phase behavior of the fluid is very similar to that of the SW fluid counterpart. More recently
\cite{Fantoni11}, we have tackled the same issue in the three-dimensional fluid. Here we build upon this work by presenting a detailed Monte Carlo study of the phase diagram for different values of penetrability and well
width. {In} this case the PSW fluid is proven to satisfy Ruelle's criterion below a well-defined value of penetrability that is essentially related to the number of interacting particles for a specific range of attractive
interaction. For higher values of penetrability, we find an \emph{intermediate} region where, {although} the system is {thermodynamically unstable (non-extensive) in the  limit $N\to\infty$}, it displays a
{``normal''} behavior, with both fluid-fluid and fluid-solid transitions,  for \emph{finite} number of particles $N$. The actual limit of this {intermediate} region depends critically upon the considered
{temperatures, densities, and size} of the system. Here the phase diagram is similar to that of the SW counterpart, although the details of the critical lines and point location depend upon the actual
penetrability value. For {even} higher penetrability, the system becomes unstable at any studied value of $N$ and the fluid evolves into clusters of overlapping particles arranged into an ordered phase at high
concentration, with a phenomenology {reminiscent of} that displayed by the PS model, {but with non-extensive properties}.

The remaining of the paper is organized as follows. In Section \ref{sec:model} we
define the PSW model and in Section \ref{sec:ruelle} we set the {conditions} for Ruelle's criterion to be valid.
The {behavior of the system outside those conditions}  is studied in Section \ref{sec:pd}, where we also determine the fluid-fluid
coexistence curves for the PSW model just below the threshold line found
before; in Section \ref{sec:nvt} we determine the instability line, in the
temperature-density plane, separating the metastable normal phase from the unstable blob phase.
Section \ref{sec:solid} is devoted to the fluid-solid transition and in Section \ref{sec:conclusions}
we draw some conclusive remarks and perspectives.
\section{The Penetrable Square-Well model}
\label{sec:model}
The PSW model is defined by the following pair potential
\begin{eqnarray}
\label{model:eq1}
\phi(r)=\left\{
\begin{array}{cl}
\epsilon_r~, & r\le\sigma~,\\
-\epsilon_a~, & \sigma<r\le\sigma+\Delta~,\\
0          ~, & r>\sigma+\Delta~,
\end{array}\right.
\end{eqnarray}
where $\epsilon_r$ and $\epsilon_a$ are two positive constants accounting for the repulsive and attractive parts of the potential, respectively, $\Delta$ is the width of the attractive square well, and $\sigma$ is
diameter of the repulsive core.

\begin{figure}
\begin{center}
\includegraphics[height=15cm]{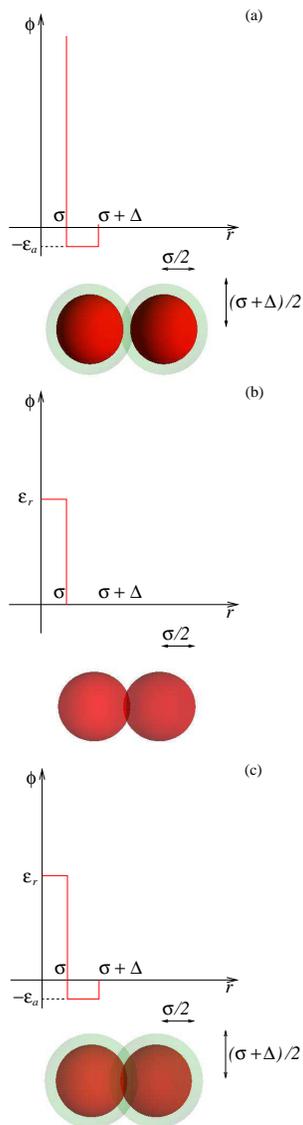}
\end{center}
\caption{Sketch of the PSW potential used in the present work (c). This potential interpolates between
the SW potential (a) and of the PS potential (b). In the SW case (a), spherical particles
have a perfect steric hindrance of size $\sigma$ (the particle diameter) and attractive interactions of range $\sigma+\Delta$ highlighted as a halo in the picture. In the PS case (b), nearest-neighbor particles can partially
interpenetrate, with some energy cost $\epsilon_r$, but have no attractive tail. In the PSW there is both the possibility of partial interpenetration (with cost $\epsilon_r$) and short-range SW attraction (with energy gain
$\epsilon_a$). } \label{fig:fig1}
\end{figure}

As discussed above, this model encompasses both the possibility of a partial interpenetration, with an energy cost typical of the soft-matter interactions given by $\epsilon_r$, and a short-range
attraction typical of both simple and complex fluids given by $\epsilon_a$. Both descriptions can be clearly recovered as limiting cases of the PSW potential: For $\epsilon_r\to\infty$ it
reduces to the SW model, while for $\Delta=0$ or $\epsilon_a=0$ one recovers the PS model \cite{Malijevsky06,Malijevsky07}. Figure \ref{fig:fig1} displays the characteristics of the PSW potential (c), along with the two
particular cases, SW (a) and PS (b). The interplay between the two energy scales $\epsilon_r$ and $\epsilon_a$ gives rise to a number of rather unusual and peculiar features that are the main topic of this paper.

In order to put the PSW model in perspective, let us briefly summarize the main features of the SW and PS potentials.

The SW model has a standard phase diagram typical of a simple fluid, with a fluid-fluid and a fluid-solid transitions in the intermediate range between the triple and the critical points in the temperature-density plane. The
fluid-fluid transition becomes metastable, against crystallization, if the width of the well goes below a certain value that has been estimated to be $\Delta \approx 0.25 \sigma$ \cite{Liu05}.

The PS fluid, on the other hand, does not display any fluid-fluid coexistence, in view of the lack of any attractive interactions. The fluid-solid transition is, however, possible and highly unconventional with the
formation of multiple occupancy crystals coupled with possible reentrant melting in the presence of a smoother repulsive interaction, such as a Gaussian form \cite{Mladek08,Klein94}.

The PSW fluid combines features belonging to both limiting cases within a very subtle interplay between the
repulsive and attractive energy scale that affects its thermodynamic stability \cite{Santos08,Fantoni09,Fantoni10a}.

\section{Ruelle's stability criterion}
\label{sec:ruelle}
The issue of thermodynamic stability has a long and venerable history, dating back to the late sixties \cite{Fisher66},
and it is nicely summarized in Ruelle's textbook that is a standard reference for this problem \cite{Ruelle69}.

A system is defined to be (Ruelle) thermodynamically stable \cite{Ruelle69,Fisher66} if
there exists a positive number $B$, such that for the total potential energy ${\Phi}_N$ for a system of $N$ particles it holds
\begin{eqnarray}
\label{ruelle:eq0} {\Phi}_N \ge - N B.
\end{eqnarray}

The physical rationale behind this mathematical statement is that the ratio $-{\Phi}_N/N$ cannot grow unboundly as $N$ increases if the system is to be well behaving, but must converge to a well defined limit.
This is usually referred to as Ruelle's stability criterion.

Consider the PSW fluid. As density increases and temperature decreases, particles tend to lump together into  clusters (``blobs'') as they pay some energetic price set by $\epsilon_r$ but they gain a
(typically larger) advantage due to the attraction $\epsilon_a$. Therefore, as the ratio
$\epsilon_a/\epsilon_r$ increases, one might expect to reach an unstable regime with very few clusters including a large number of significantly overlapping particles, so that ${\Phi}_N$ is no longer proportional to $N$.

The ratio $\epsilon_a/\epsilon_r$ (``penetrability'') plays in PSW fluids a very important role, as we shall see in the following sections.
In Ref.\ \cite{Fantoni10a} we proved that the {one-dimensional (1D)} PSW fluid satisfies Ruelle's criterion if $\epsilon_a/\epsilon_r \leq 1/2(\ell + 1)$, where $\ell$ is
the integer part of $\Delta/\sigma$. In this case, we are then guaranteed to have a well defined equilibrium state.

Here we show that this result can be extended to a three-dimensional (3D) PSW  fluid in that Ruelle's criterion is satisfied if $\epsilon_a/\epsilon_r\leq 1/f_{\Delta}$, where $f_{\Delta}$ is the {maximum number
of non-overlapping particles that can be geometrically arranged around a given one within a distance between $\sigma$ and $\sigma+\Delta$. Of course, $f_\Delta$} depends on the width of the attractive interaction $\Delta$. For $\Delta/\sigma < \sqrt{2}-1$, for instance, one has  $f_{\Delta}=12$,
corresponding to a HCP closed packed configuration. In the following, we will use a generic $d$-dimensional notation and consider $d=3$ at the end.

The total potential energy of a PSW fluid formed by particles at positions $\mathbf{r}_{1},\ldots,\mathbf{r}_{N}$ can be written in general as
\begin{eqnarray}
\label{ruelle:eq1}
{\Phi}_N\left(\mathbf{r}_{1},\ldots,\mathbf{r}_{N} \right) =\frac{1}{2}  \sum_{i=1}^{N} \sum_{j\ne i}^{N} \phi\left(\left \vert \mathbf{r}_{i}
- \mathbf{r}_{j} \right \vert \right)
\end{eqnarray}

Consider now such a configuration where particles are distributed in $M$ clusters along each direction, each made of $s$ perfectly overlapped particles, and with different clusters {arranged} in the close-packed configuration. In
the Appendix  we prove that indeed this is the lowest possible energy configuration {in the two-dimensional (2D) case}.

The total number of particles is $N=M^d s$. As clusters are in a close-packed
configuration, particles of a given cluster interact
attractively with all the particles of those $f_\Delta$ clusters
within a distance smaller than $\sigma+\Delta$. {Consequently, the potential energy has the form}
\begin{eqnarray}
\label{ruelle:eq3}
{{\Phi}_N(M) = \frac{1}{2} M^d s \left(s-1\right)\epsilon_r- \frac{M^d}{2}\left[
  {f_\Delta}-b_\Delta(M)\right]s^2\epsilon_a.}
\end{eqnarray}
The first term represents  the repulsive energy between all possible pairs of particles in a given $s$-cluster,
while the second term represents the attractive energy between clusters.
Here $b_\Delta(M)$ accounts for a reduction of the actual
number of clusters interacting attractively, due to boundary
effects. This quantity clearly depends upon the chosen value of $\Delta/\sigma$ but we can infer the following general properties
\begin{eqnarray}
\label{ruelle:eq4}
{b_\Delta(1)=f_\Delta, \qquad b_\Delta(M>2)< f_\Delta , \qquad \lim_{M\to\infty}b_\Delta(M)=0.}
\end{eqnarray}
In the 1D  (with $\Delta/\sigma<1$) and 2D  (with
$\Delta/\sigma<\sqrt{3}-1$) cases, ${\Phi}_N(M)$ is given by Eqs.\ \eqref{app:eq1} and \eqref{app:eq8}, respectively, so that
{$b_\Delta(M)=2M^{-1}$} (1D) and {$b_\Delta(M)=2(4M^{-1}-M^{-2})$} (2D). In general, $b_\Delta(M)$ must be  a positive definite
polynomial of degree  {$d$ in $M^{-1}$ with no independent term}, its form becoming more complicated as $d$
increases. However, we need not specify the actual form
of $b_\Delta(M)$ for our argument, but only the properties given in Eq.\ (\ref{ruelle:eq4}).

Eliminating $s=N/M^d$ in favor of $M$ in Eq.\ (\ref{ruelle:eq3}) one easily gets
\begin{eqnarray}
\label{ruelle:eq5}
\frac{{\Phi}_N(M)}{N}=-\frac{\epsilon_r}{2}+\frac{N}{2}\epsilon_a M^{-d}F(M),
\end{eqnarray}
where we have introduced the function
\begin{eqnarray}
\label{ruelle:eq6}
{F(M)\equiv b_\Delta(M)-\left(f_\Delta-\frac{\epsilon_r}{\epsilon_a}\right).}
\end{eqnarray}
{Note that $F(M)$ is independent of $N$.} If ${\epsilon_a}/{\epsilon_r} <1/f_\Delta$, $F(M)$ is positive definite and so
${\Phi}_N/N$ has a lower bound {($-\epsilon_r/2$)} and the system is stable in the
thermodynamic limit. Let us suppose now that
${\epsilon_a}/{\epsilon_r}>1/f_\Delta$. In that case,
$F(1)={\epsilon_r}/{\epsilon_a}>0$ but
$\lim_{M\to\infty}F(M)=-(f_\Delta-{\epsilon_r}/{\epsilon_a})<0$. Therefore,
there must exist a certain finite value $M=M_0$ such that $F(M)<0$ for
$M>M_0$. In the 1D (with $\Delta/\sigma<1$) and 2D  (with
$\Delta/\sigma<\sqrt{3}-1$) cases the values of $M_0$ can be explicitly computed:
\begin{eqnarray}
\label{ruelle:eq7}
M_0&=&\left(1-\frac{\epsilon_r}{2\epsilon_a}\right)^{-1},\quad (d=1),
\end{eqnarray}
\begin{eqnarray}
\label{ruelle:eq8}
M_0 &=&\frac{2+\sqrt{1+{\epsilon_r}/{2\epsilon_a}}}{3}
\left(1-\frac{\epsilon_r}{6\epsilon_a}\right)^{-1},\quad (d=2).
\end{eqnarray}
In general, it is reasonable to expect that $M_0\sim
\left(1-{\epsilon_r}/{f_\Delta\epsilon_a}\right)^{-1}$. {Regardless of the precise value of $M_0$, we have that $\lim_{N\to\infty}[-{\Phi}_N(M)]/N=\infty$ for $M>M_0$ and thus the criterion \eqref{ruelle:eq0} is violated. }

This completes the proof that, if ${\epsilon_a}/{\epsilon_r}<1/f_\Delta$, the system is thermodynamically stable as it satisfies Ruelle's stability
criterion, Eq.\ (\ref{ruelle:eq0}). {Reciprocally, if ${\epsilon_a}/{\epsilon_r}>1/f_\Delta$ there exists a class of blob configurations violating Eq.\ \eqref{ruelle:eq0}. In those configurations the $N$ particles are concentrated on a finite (i.e., independent of $N$) number of clusters, each with a number of particles proportional to $N$. For large $N$ the potential energy scales with $N^2$ and thus the system exhibits non-extensive properties.}

{In three dimensions, $f_\Delta=12$, $18$, and $42$ if $\Delta/\sigma<\sqrt{2}-1$,
$\sqrt{2}-1<\Delta/\sigma<\sqrt{3}-1$, and
$\sqrt{3}-1<\Delta/\sigma<1$, respectively, and so the threshold values are
${\epsilon_a}/{\epsilon_r}=1/12$, $1/18$, and $1/42$, respectively}. There might (and do) exist local configurations with
higher coordination numbers, but only those filling the whole space
have to be considered in the thermodynamic limit.

{In general, Ruelle's criterion \eqref{ruelle:eq0} is a sufficient but not necessary condition for thermodynamic stability. Therefore, in principle, if $\epsilon_a/\epsilon_r >  1/f_{\Delta}$
the system may or may not be stable, depending on the physical state (density $\rho$ and temperature $T$). However, compelling arguments discussed in Ref.\ \cite{Ruelle69} show that the PSW system with $\epsilon_a/\epsilon_r >  1/f_{\Delta}$ is indeed unstable (i.e., non-extensive) in the thermodynamic limit for any $\rho$ and $T$. Notwithstanding this, even if $\epsilon_a/\epsilon_r >  1/f_{\Delta}$, the system may exhibit ``normal'' (i.e., extensive) properties at \emph{finite} $N$, provided the temperature is sufficiently high and/or the density is sufficiently low.} It is therefore interesting to investigate this regime
with the specific goals of (i) defining the stability boundary (if any) and (ii) outlining the fate of the {SW-like}  fluid-fluid and fluid-solid
lines as penetrability increases. This will be discussed in the next section, starting from the fluid-fluid coexistence lines.

\section{Effect of penetrability on the fluid-fluid coexistence}
\label{sec:pd}
We have performed an extensive analysis of the fluid-fluid phase transition of the three-dimensional PSW fluid by Gibbs Ensemble Monte Carlo (GEMC) simulations
\cite{Frenkel96,Panagiotopoulos87,Panagiotopoulos88,Smit89a,Smit89b}. In all cases we have started with the SW fluid ($\epsilon_a/\epsilon_r=0$) and gradually increased penetrability  $\epsilon_a/\epsilon_r$ until
disappearance of the transition. Following standard prescriptions \cite{Frenkel96,Panagiotopoulos87,Panagiotopoulos88,Smit89a,Smit89b}, we construct the fluid-fluid coexistence lines using two systems (the gas and the liquid)
that can exchange both volume and particles in such a way that the total volume $V$ and the total number of particles $N$ are fixed {and the pressure and chemical potential coincide in both systems}. {$N=512$ particles
were used. By denoting with $L_{i}$ and $V_{i}$ ($i=v,l$) the respective sizes and volumes of the vapor and liquid boxes, we used $2N$ particle random displacements of magnitude $0.15 L_i$, $N/10$ random volume changes of
magnitude $0.1$ in $\ln[V_i/(V-V_i)]$, and $N$ particle swaps between the gas and the liquid boxes, on average per cycle.}

Our code fully reproduces the results of Vega {\etal} \cite{Vega92} for the SW fluid, as further discussed below.
\begin{figure}
\begin{center}
\includegraphics[width=7cm]{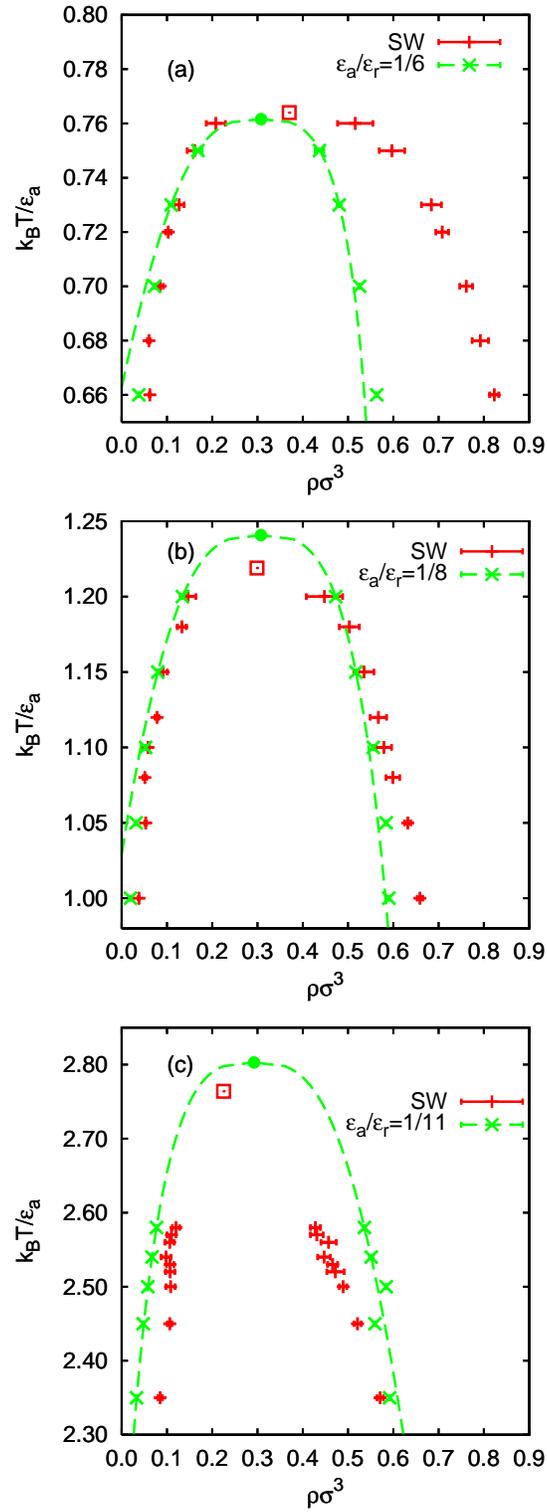}
\end{center}
\caption{Fluid-fluid coexistence lines for different well widths $\Delta/\sigma$
and penetrabilities $\epsilon_a/\epsilon_r$. The SW results are
those by Vega {\etal} \cite{Vega92} for the same value of
$\Delta/\sigma$. Circles and boxes represent the estimated critical
points for the PSW and the SW fluids, respectively, and the dotted lines represent the
coexistence curves for the PSW case. (a) $\Delta/\sigma=0.25$ and
$\epsilon_a/\epsilon_r=1/6$; (b) $\Delta/\sigma=0.5$ and $\epsilon_a/\epsilon_r=1/8$;
(c) $\Delta/\sigma=1$ and $\epsilon_a/\epsilon_r=1/11$.}
\label{fig:fig2}
\end{figure}
Figure \ref{fig:fig2} depicts some representative examples of the effect of penetrability on the SW results
at different well widths $\Delta/\sigma$. As $\Delta/\sigma$ increases, the upper limit set by Ruelle's stability condition
$\epsilon_a/\epsilon_r\leq 1/f_{\Delta}$ decreases, and lower penetrability values $\epsilon_a/\epsilon_r$
have to be used to ensure the existence of the transition line. In Fig.\ \ref{fig:fig2}, values
$\epsilon_a/\epsilon_r=1/6,1/8,1/11$ were used for $\Delta/\sigma=0.25,0.5,1$, respectively.
Figure \ref{fig:fig2} also includes an estimate of the critical points
for the PSW fluid obtained from the law of rectilinear diameters, as discussed in Ref.\ \cite{Vega92}, that is
\begin{eqnarray}
\label{coexistence:eq1}
\frac{\rho_l+\rho_v}{2}=\rho_c+A(T_c-T)~,
\end{eqnarray}
where $\rho_l$ ($\rho_v$) is the density of the liquid (vapor) phase,
$\rho_c$ the critical density and $T_c$ the critical
temperature. Furthermore, the temperature dependence of the density
difference of the coexisting phases is fitted to the following scaling form
\begin{eqnarray}
\label{coexistence:eq2}
\rho_l-\rho_g=B(T_c-T)^\beta~,
\end{eqnarray}
where  the critical exponent for the three-dimensional
Ising model $\beta=0.32$ was used to match the universal fluctuations. Amplitudes $A$ and $B$ where determined from the fit.

A detailed collection of the results corresponding to Fig.\ \ref{fig:fig2}(a), (b), and (c) is
reported in Table \ref{tab:1}.
\begin{table}
\begin{center}
\begin{tabular}{ccccccc}
\multicolumn{7}{c}{$\Delta/\sigma=0.25$, $\epsilon_a/\epsilon_r=1/6$}\\
\hline
$k_BT/\epsilon_a$ & $\rho_v\sigma^3$ &
$\rho_l\sigma^3$ & $u_v/\epsilon_a$ & $u_l/\epsilon_a$ & $\mu_v-k_BT\ln\Lambda^3$ & $\mu_l-k_BT\ln\Lambda^3$\\
\hline
$0.66$  &$0.0377(6)$    &$0.5634(6)$&$-0.343(8)$&$-3.441(13)$&$-2.410(7)$&$-2.51(12)$\\
$0.70$  &$0.0724(15)$&$0.5256(15)$&$-0.614(16)$&$-3.100(13)$&$-2.253(5)$&$-2.27(6)$\\
$0.73   $&$0.1093(45)$&$0.4805(42)$&$-0.862(38)$&$-2.920(45)$&$-2.157(12)$&$-2.29(8)$\\
$0.75   $&$0.1684(95)$&$0.4368(95)$&$-1.204(67)$&$-2.682(27)$&$-2.211(8)$&$-2.01(2)$\\
\hline
\end{tabular}

\vspace{0.5cm}

\begin{tabular}{ccccccc}
\multicolumn{7}{c}{$\Delta/\sigma=0.5$, $\epsilon_a/\epsilon_r=1/8$}\\
\hline
$k_BT/\epsilon_a$ & $\rho_v\sigma^3$ &
$\rho_l\sigma^3$ & $u_v/\epsilon_a$ & $u_l/\epsilon_a$ & $\mu_v-k_BT\ln\Lambda^3$ & $\mu_l-k_BT\ln\Lambda^3$\\
\hline
$1.00   $&$0.0194(4)$&$0.5900(7)$&$-0.254(7)$&$-4.687(9)$&$-4.19(2)$&$-4.16(5)$\\
$1.05   $&$0.0319(5)$&$0.5841(17)$&$-0.400(9)$&$-4.603(14)$&$-4.00(1)$&$-4.01(3)$\\
$1.10   $&$0.0529(8)$&$0.5557(8)$&$-0.651(14)$&$-4.365(6)$&$-3.832(6)$&$-3.83(4)$\\
$1.15   $&$0.0799(15)$&$0.5173(17)$&$-0.934(18)$&$-4.087(15)$&$-3.726(7)$&$-3.76(4)$\\
$1.20   $&$0.1342(37)$&$0.4728(40)$&$-1.464(40)$&$-3.777(26)$&$-3.642(6)$&$-3.64(2)$\\
\hline
\end{tabular}

\vspace{0.5cm}

\begin{tabular}{ccccccc}
\multicolumn{7}{c}{$\Delta/\sigma=1.0$, $\epsilon_a/\epsilon_r=1/11$}\\
\hline
$k_BT/\epsilon_a$ & $\rho_v\sigma^3$ &
$\rho_l\sigma^3$ & $u_v/\epsilon_a$ & $u_l/\epsilon_a$ & $\mu_v-k_BT\ln\Lambda^3$ & $\mu_l-k_BT\ln\Lambda^3$\\
\hline
$2.35   $&$0.0327(4)$&$0.5920(11)$&$-0.693(8)$&$-8.931(12)$&$-8.90(2)$&$-8.87(6)$\\
$2.45   $&$0.0476(5)$&$0.5593(16)$&$-1.004(11)$&$-8.439(21)$&$-8.66(1)$&$-8.61(3)$\\
$2.50   $&$0.0577(8)$&$0.5844(12)$&$-1.201(17)$&$-8.653(17)$&$-8.54(2)$&$-8.59(5)$\\
$2.54   $&$0.0670(12)$&$0.5511(37)$&$-1.377(25)$&$-8.231(42)$&$-8.48(2)$&$-8.51(2)$\\
$2.58   $&$0.0769(9)$&$0.5361(19)$&$-1.556(20)$&$-8.030(22)$&$-8.41(1)$&$-8.38(3)$\\
\hline
\end{tabular}
\caption{{Vapor-liquid coexistence data from GEMC of $N=512$ PSW
particles with $\Delta/\sigma=0.25$ and $\epsilon_a/\epsilon_r=1/6$ (top table), $\Delta/\sigma=0.5$ and $\epsilon_a/\epsilon_r=1/8$
(central table) and  $\Delta/\sigma=1.0$ and $\epsilon_a/\epsilon_r=1/11$ (bottom table). 
We used $10^7$ MC  steps. $T$, $\rho_i$, $u_i$, $\mu_i$ are,
respectively, the temperature, the density, the internal energy per
particle, and the chemical potential of the vapor ($i=v$) or liquid
($i=l$) phase ($\Lambda$ being the thermal de Broglie wavelength). Numbers in parentheses correspond to the error on the last
digits. The estimated critical points are $k_BT_c/\epsilon_a=0.762$ and $\rho_c\sigma^3=0.307$.(top table),
$k_BT_c/\epsilon_a=1.241$ and $\rho_c\sigma^3=0.307$ (central table) and $k_BT_c/\epsilon_a=2.803$ and $\rho_c\sigma^3=0.292$ (bottom table) }}
\label{tab:1}
\end{center}
\end{table}

\begin{figure}
\begin{center}
\includegraphics[height=10cm]{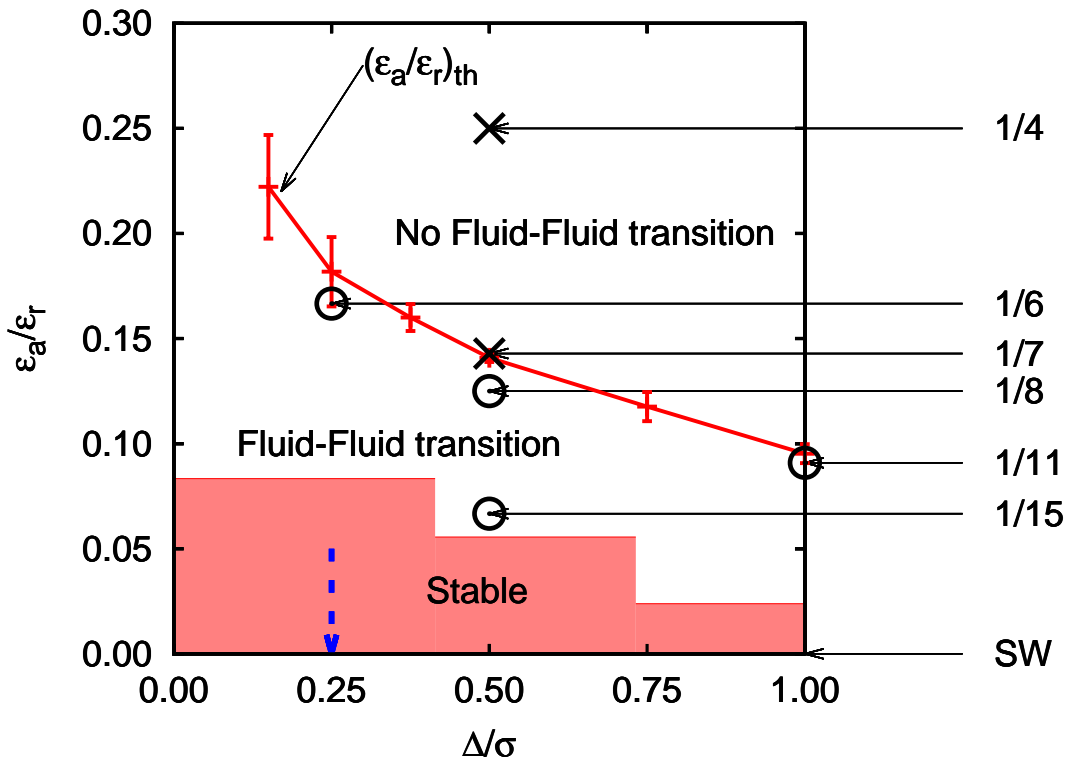}
\end{center}
\caption{Plot of penetrability
$\epsilon_a/\epsilon_r$ as a function of $\Delta/\sigma$. The
displayed $(\epsilon_a/\epsilon_r)_{\text{th}}$ line separates the parameter region where the PSW model,
with $N=512$, admits a fluid-fluid phase transition (below this line) from that
where it does not. The shadowed stepwise line highlights the region
($\epsilon_a/\epsilon_r\leq 1/12$ for $\Delta/\sigma<\sqrt{2}-1$,
$\epsilon_a/\epsilon_r\leq 1/18$ for
$\sqrt{2}-1<\Delta/\sigma<\sqrt{3}-1$, and $\epsilon_a/\epsilon_r\leq
1/42$ for $\sqrt{3}-1<\Delta/\sigma<1$) where the model is guaranteed to be
thermodynamically stable  for any thermodynamic state by Ruelle's criterion. The SW model falls on the
$\epsilon_a/\epsilon_r =0 $ axis (with
finite $k_BT/\epsilon_a$). The vertical dashed arrow points to the SW value $\Delta/\sigma\lesssim 0.25$ below which
the fluid-fluid transition becomes metastable against the freezing transition  \cite{Liu05}. The circles are the points chosen for the calculation
of the coexistence lines (Figs.\ \ref{fig:fig2} and \ref{fig:fig9}), while the crosses are the points chosen for the
determination of the boundary phases discussed in Figs.\ \ref{fig:fig5} and  \ref{fig:fig6}.}
\label{fig:fig3}
\end{figure}

Note that seemingly stable transition curves are found in all representative cases depicted in Fig. \ref{fig:fig2}, thus suggesting {a ``normal'' fluid behavior for the finite-size system studied}. Increasing penetrability  $\epsilon_a/\epsilon_r$ at fixed $\Delta/\sigma$
progressively destabilize the transition, until a threshold value  $(\epsilon_a/\epsilon_r)_{\text{th}}$ is reached where no fluid-fluid
transition is observed. Upon changing $\Delta/\sigma$, one can then draw a line of this values in the $\epsilon_a/\epsilon_r$ and
$\Delta/\sigma$ plane. This is depicted in Fig. \ref{fig:fig3}, where the instability line $(\epsilon_a/\epsilon_r)_{\text{th}}$
is found to decrease as $\Delta/\sigma$ increases, thus gradually reducing the region where the fluid-fluid transition
can be observed, as expected. The shadowed stepwise region identifies the thermodynamically stable region, as guaranteed by Ruelle's criterion
$\epsilon_a/\epsilon_r\leq 1/f_{\Delta}$ discussed above.
Note that points $(\Delta/\sigma=0.25,\epsilon_a/\epsilon_r=1/6)$, $(\Delta/\sigma=0.5,\epsilon_a/\epsilon_r=1/8)$, and
$(\Delta/\sigma=1,\epsilon_a/\epsilon_r=1/11$), corresponding to the values used in Fig.\ \ref{fig:fig2} and highlighted by circles,
lie in the  $1/f_{\Delta} \leq \epsilon_a/\epsilon_r \leq (\epsilon_a/\epsilon_r)_{\text{th}}$ region, that is,
outside the stable range guaranteed by Ruelle's criterion.

\section{Stable, unstable, and metastable phases}
\label{sec:nvt}
Interestingly, in Ruelle's textbook \cite{Ruelle69},
the three-dimensional PSW model corresponding to point $(\Delta/\sigma=1,\epsilon_a/\epsilon_r=1/11$)
is exploited  as an example of ``catastrophic'' fluid (see especially Fig.\ 4
and proposition 3.2.2 both in Ref.\ \cite{Ruelle69}). This is clearly because this state point lies
outside the stable region identified by Ruelle's criterion, as discussed.
As already remarked, however, this criterion does not necessarily imply that outside this
region the system has to be unstable, but only that it is ``likely'' to be so. There are then two possibilities. First, that in the intermediate
region  $1/f_{\Delta} \leq \epsilon_a/\epsilon_r \leq (\epsilon_a/\epsilon_r)_{\text{th}}$ the system
is indeed stable in the thermodynamic limit, a case that is not covered by Ruelle's criterion. Numerical results reported in Figs.\ \ref{fig:fig2} and
\ref{fig:fig3} appear to support this possibility. The second possibility is that, even in this region, the system is strictly unstable, in the thermodynamic
limit, but it appears to be a ``normal'' fluid when considered at finite $N$. This possibility cannot be ruled out by any simulation at finite $N$,
and would be more plausible as hinted by Ruelle's arguments.

\begin{figure}
\begin{center}
\includegraphics[height=12cm]{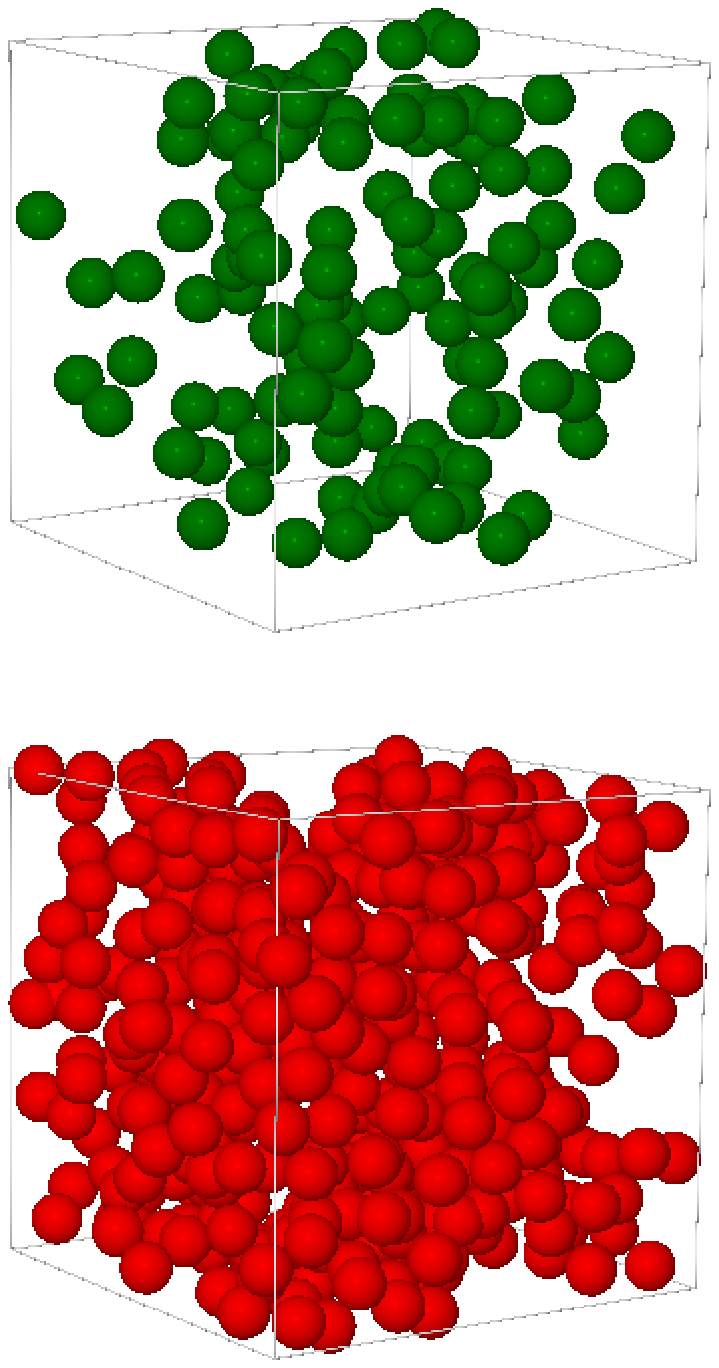}
\end{center}
\caption{Two GEMC simulation snapshots ($N=512$) at $\Delta/\sigma=0.5$,
$\epsilon_a/\epsilon_r=1/8$ (below the {threshold} value) and
$k_BT/\epsilon_a=1.20$. The one on the top panel
corresponds to the gas phase {($\rho_v\sigma^3=0.1342$)}, and the one on the bottom to the
liquid phase {($\rho_l\sigma^3=0.4728$)}.}
\label{fig:fig4}
\end{figure}

In order to illustrate the fact that, at finite $N$, the system in the intermediate region
$1/f_{\Delta} \leq \epsilon_a/\epsilon_r \leq (\epsilon_a/\epsilon_r)_{\text{th}}$ behaves as a normal fluid, in Fig.\ \ref{fig:fig4} we show two representative snapshots
of the gas and the liquid phases at the point $(\Delta/\sigma=0.5,\epsilon_a/\epsilon_r=1/8)$ that lies just below the $(\epsilon_a/\epsilon_r)_{\text{th}}$
line (see Fig.\ \ref{fig:fig3}). In both the gas and the liquid phases, the structure of the fluid presents the typical features of a standard SW fluid, with no significant overlap among different particles.

On the other hand, we have observed that above the threshold line $(\epsilon_a/\epsilon_r)_{\text{th}}$
of Fig.\ \ref{fig:fig3}, at a temperature close to
the critical temperature of the corresponding SW system, the GEMC simulation evolves towards an empty box and a collapsed
configuration in the liquid box.

The second scenario described above can be supported or disproved by a finite-size study of the $N$-dependence of the
transition, as described below.

Assume that at any finite $N$, the absolute minimum of the internal energy corresponds to the
``collapsed'' non-extensive configurations, referred to as ``blob phase'' in the following. {As discussed in section \ref{sec:ruelle}, the internal energy of these configurations scales with $N^2$ for large $N$. However, the system presumably
also includes a large number of ``normal'' configurations with an internal energy that scales linearly
with $N$. This will be referred to as ``normal phase''.

There is then an energy gap between the total energy associated
with the normal and the collapsed configurations with an energy
ratio of order $N$. For finite $N$ and sufficiently high temperature, the {Boltzmann statistical factor
$e^{-{\Phi}_N/k_BT}$ of the collapsed configurations (in spite of the gap) might be not sufficiently large 
to compensate for the fact that the volume in phase space
corresponding to normal configurations has a much larger measure (and hence entropy) than that corresponding to collapsed configurations.} 
As a consequence, the
physical properties look like normal and one observes a normal phase. Normal configurations have a higher internal energy but also {may have} a larger entropy. If $N$ is sufficiently small and/or $T$ is high enough, normal
configurations might then have a smaller free energy than collapsed configurations. {On the other hand,}  the situation is reversed at larger $N$ and finite temperature, where the statistical weight {(i.e., the interplay
between the Boltzmann factor and the measure of the phase space volume)} of the collapsed configurations becomes comparable to (or even larger than) that of the normal configurations and physical properties become
anomalous. This effect could be avoided only if $T$ grows (roughly proportional to $N$) as $N$ increases, since entropy increases more
slowly with $N$ than $\Phi_N$. 

\begin{figure}
\begin{center}
\includegraphics[width=12cm]{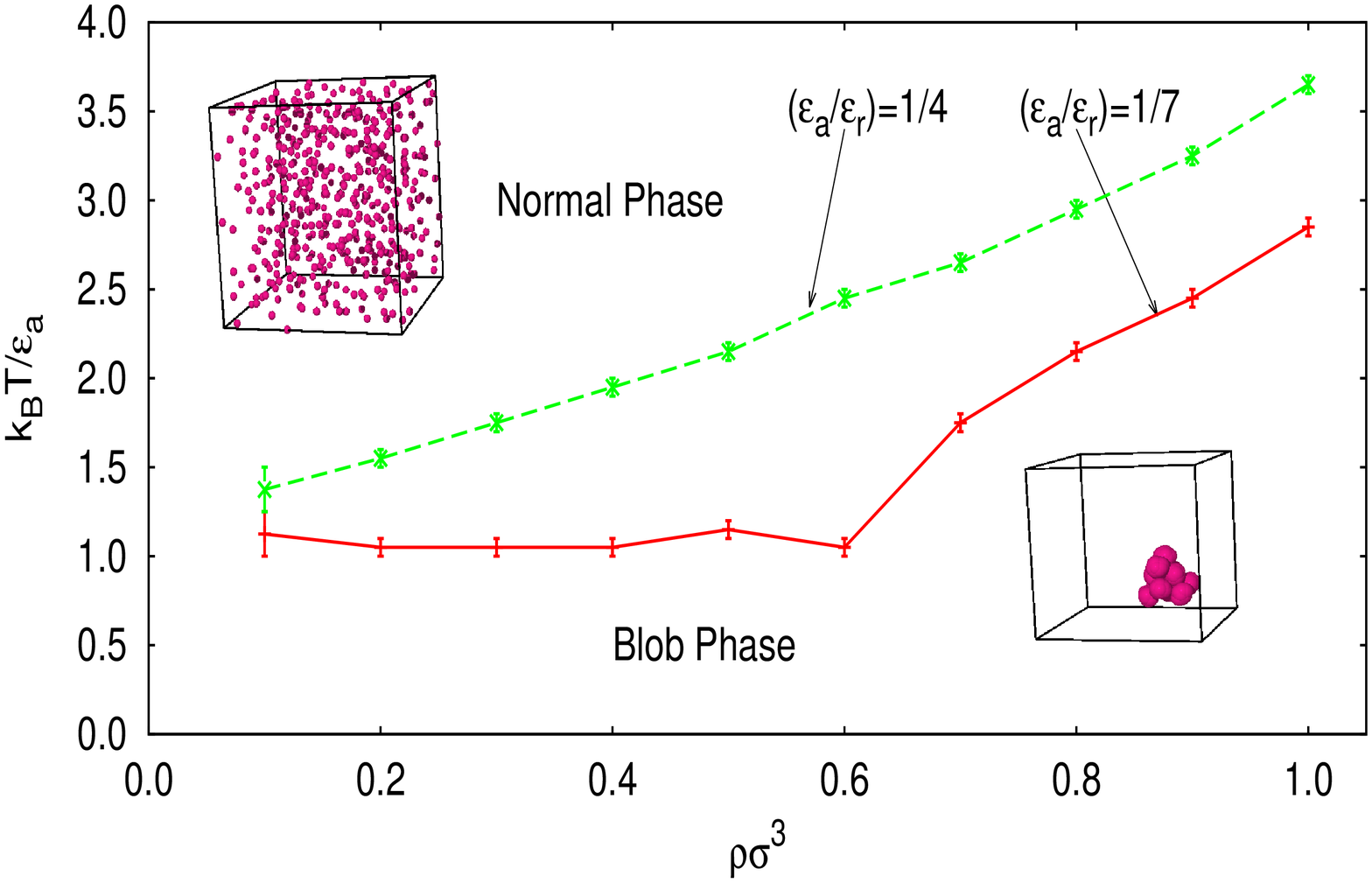}
\end{center}
\caption{Regions of the phase diagram where the PSW fluid,
with $\Delta/\sigma=0.5$ and two different values of
$\epsilon_a/\epsilon_r$, is expected to {exhibit a normal phase (above the instability line) or a blob phase (below the instability line) for  $N=512$ particles}.  Note that the instability line  corresponding to
the higher penetrability case ($\epsilon_a/\epsilon_r=1/4$, dashed line) lies above the one
corresponding to the lower penetrability ($\epsilon_a/\epsilon_r=1/7$, solid line). The two insets
depict representative snapshots of respective typical configurations.}
\label{fig:fig5}
\end{figure}
In a PSW fluid above the stable region ($\epsilon_a/\epsilon_r > 1/f_{\Delta}$), we have then to discriminate whether the system is truly  stable in the thermodynamic  limit
$N \to\infty$, or it is metastable, evolving into an unstable blob phase  at a given {value of $N$} depending on temperature and density. 

In order to shed some more light into this dual metastable/unstable
scenario, we performed NVT Monte Carlo simulations using $N=512$ particles initially distributed uniformly within the simulation box (``regular'' initial condition). We carefully monitored the total potential energy of the
fluid during the simulation and found that, at any given density, there exists a certain temperature $T_\text{ins}(\rho)$, such that the system behaves normally after $10^7 N$ single particle moves (normal phase) if  $T >
T_\text{ins}$ and collapses to a few { clusters of overlapped particles} (blob phase) for $T < T_\text{ins}$.

This is shown in Fig.\ \ref{fig:fig5} {for $\Delta/\sigma=0.5$} and two different penetrability values: $\epsilon_a/\epsilon_r=1/4$ (upper dashed line)
and $\epsilon_a/\epsilon_r=1/7$ (lower solid line). The first value lies deeply in the instability region
above the threshold $(\epsilon_a/\epsilon_r)_{\text{th}}$ value of Fig.\ \ref{fig:fig3}, while the second
is sitting right on its top, for this value $\Delta/\sigma=0.5$ of the well width.
Also depicted are two snapshots of two representative configurations found under these conditions. {While the particles in the normal phase, $T>T_\text{ins}$, are arranged in a disordered  configuration that spans the whole box  (see upper snapshot of Fig.\ \ref{fig:fig5}),}
one can clearly see that for $T < T_\text{ins}$ a ``blob'' structure has nucleated around a certain
point within the simulation box with a few droplets of several particles
each (see lower snapshot of Fig.\ \ref{fig:fig5}).

The three fluid-fluid coexistence phase diagrams displayed in Fig.\ \ref{fig:fig2} are then representative of a metastable normal phase that persists, for a given $N$, up to  $(\epsilon_a/\epsilon_r)_{\text{th}}$ as
long as the corresponding critical point $(\rho_c,T_c)$ is such that $T_c>T_{\text{ins}}(\rho_c)$, as in the cases reported in Fig.\ \ref{fig:fig2}. Below this instability line, the fluid becomes unstable at any density and a
blob phase, where a few large clusters nucleate around certain points and occupy only a part of the simulation, is found. The number of clusters decreases (and the number of particles per cluster increases) as one moves away
from the boundary line found in Fig.\ \ref{fig:fig5} towards lower temperatures. Here a cluster is defined topologically as follows. Two particles belong to the same cluster if there is a path connecting them, where we are
allowed to move on a path going from one particle to another if the centers of the two particles are at a distance less than $\sigma$.

These results, while not definitive, are strongly suggestive of the fact that even the normal phase is in fact metastable and 
becomes eventually unstable in the $N \to \infty$ limit.

This can be further supported by a finite size scaling analysis at increasing $N$, as reported in Fig.\
\ref{fig:fig6} in the higher penetrability (and hence most demanding) case $ \epsilon_a/\epsilon_r=1/4$.
\begin{figure}
\begin{center}
\includegraphics[height=10cm]{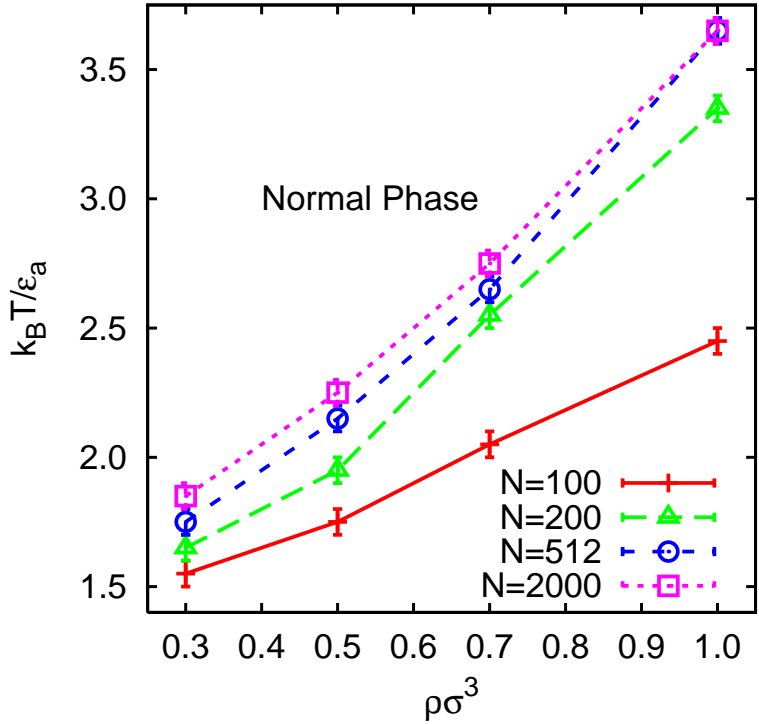}
\end{center}
\caption{Size dependence of the instability line of
Fig.\ \protect\ref{fig:fig5} for the system $\epsilon_a/\epsilon_r=1/4$ and
$\Delta/\sigma=0.5$.}
\label{fig:fig6}
\end{figure}
In obtaining these results, we used NVT
simulations with $10^{10}$ single particle moves in all cases.

As expected, the instability temperature line $T_\text{ins}(\rho)$ moves to
higher values as $N$ increases, at fixed density $\rho \sigma^3$, from {$N=100$} to $N=2000$, and  the normal
phase region significantly shrinks accordingly, being expected to vanish in the thermodynamic limit $N \to \infty$.

As said before, in all the above computations we started with a regular initial condition having all particles randomly distributed in the entire available simulation box. Under these circumstances, {for $T<T_{\text{ins}}$
(where all particles are confined into a blob of a few clusters)} a large number of MC steps is required in order to find the true equilibrium distribution. On the other hand, if we have a clustered configuration to
start with,  a much higher ``melting'' temperature {$T_{\text{ins}}$}, above which one recovers a normal phase, is expected. This ``hysteresis'' effect is indeed observed, as detailed below.

For  $\epsilon_a/\epsilon_r=1/7$,
$\Delta/\sigma=0.5$, and $\rho\sigma^3=1.0$ the normal-to-blob transition occurs  upon cooling
at $k_BT/\epsilon_a \approx 2.75$. Inserting the obtained configuration back in the MC simulation
as an initial condition, and increasing the temperature, we find the blob phase to persist up to
much higher temperatures $k_BT/\epsilon_a \approx 4$. The
hysteresis is also found to be strongly size dependent. {With the same  system $\epsilon_a/\epsilon_r=1/7$,
$\Delta/\sigma=0.5$, but for
$\rho\sigma^3=0.6$, we found  the blob-to-normal} melting temperatures to be $k_BT/\epsilon_a=2$--$3$
for $N=256$,  $k_BT/\epsilon_a=4$--$5$ for $N=512$, and
$k_BT/\epsilon_a=6$-$7$ for $N=1024$. {Analogously, in the state $\epsilon_a/\epsilon_r=1/4$,
$\Delta/\sigma=0.5$, and
$\rho\sigma^3=0.3$, the results are $k_BT/\epsilon_a=2.1$--$2.2$, $k_BT/\epsilon_a=3.7$--$3.8$, $k_BT/\epsilon_a=9.0$--$9.1$, and $k_BT/\epsilon_a=31$--$32$ for $N=100$, $N=200$, $N=512$, and $N=2000$, respectively.}

In the interpretation of the size dependence of the hysteresis in the melting, one should also consider the fact that the blob occupies only part of the simulation box and therefore a surface term has a rather high impact on the melting temperature.

Additional insights on the sudden {structural} change occurring on the {fluid} upon crossing
the threshold line $(\epsilon_a/\epsilon_r)_{\text{th}}$ can be obtained by considering the radial distribution
function (RDF) $g(r)$ \cite{tenWolde96} on two state points above and below this line.
We consider a state point at
$\Delta/\sigma=0.5$, $k_BT/\epsilon_a=1.20$, and
$\rho\sigma^3=0.7$ and evaluate the RDF at $\epsilon_a/\epsilon_r=1/8$ (slightly below
the threshold line, see Fig.\ \ref{fig:fig3})
and at $\epsilon_a/\epsilon_r=1/7$. The latter case is sitting right on the top
of the threshold line, according to Fig.\ \ref{fig:fig3}. The results are depicted
in Fig.\ \ref{fig:fig7}.

\begin{figure}
\begin{center}
\includegraphics[height=10cm]{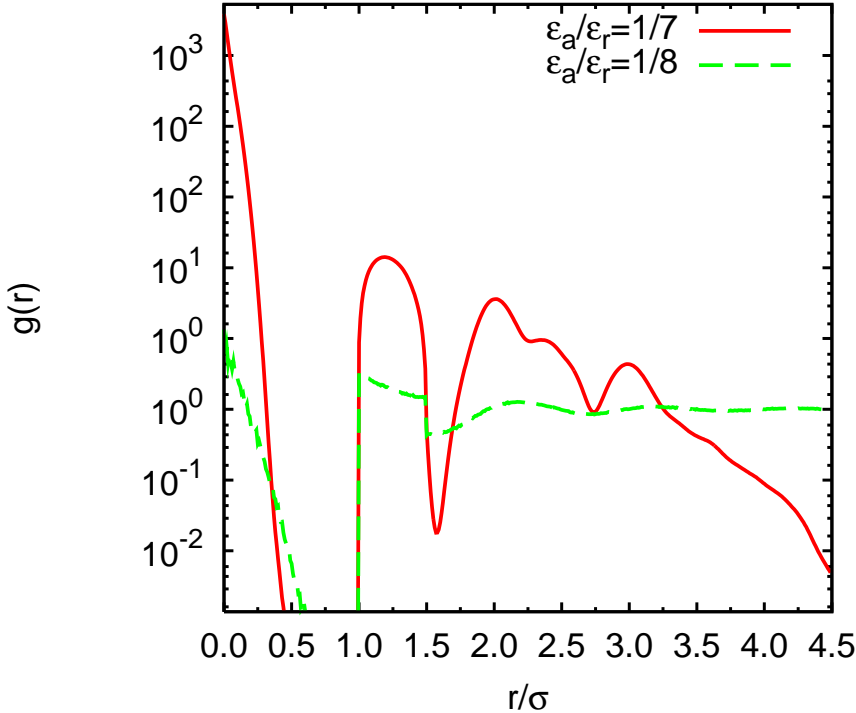}
\end{center}
\caption{Radial distribution function for the PSW model at
$\Delta/\sigma=0.5$, $k_BT/\epsilon_a=1.20$, and $\rho\sigma^3=0.7$
for two different values of the penetrability parameter
$\epsilon_a/\epsilon_r$: $\epsilon_a/\epsilon_r=1/8$ (lying below the
threshold line given in Fig.\ \protect\ref{fig:fig3}) and $\epsilon_a/\epsilon_r=1/7$ (that is
on the top of it). The $g(r)$ axis is in a log scale.}
\label{fig:fig7}
\end{figure}
Drastic changes in the structural properties of the PSW liquid are clearly noticeable. While in the normal phase ($\epsilon_a/\epsilon_r=1/8$) the RDF presents the typical features of a standard fluid for a soft-potential
and, in particular, converges to unity, in the blob phase ($\epsilon_a/\epsilon_r=1/7$), the RDF presents a huge peak (note the log-scale) at $r=0$ and decays to zero after the first few peaks, a behavior that is suggestive of clustering and confinement of
the system. The amplitude of the first maximum in the structure factor grows past the value of $2.85$, which is typically reckoned for an indication for a freezing occurring in the system, according to Ref.\
\cite{Hansen69}.

As a further characterization of the structural ordering of the system, we have
also investigated a set of rotationally invariant local order indicators that
have been often exploited to quantify order in crystalline solids,
liquids, and colloidal gels \cite{tenWolde96}:
\begin{eqnarray}
\label{stable:eq1}
Q_{l} =\sqrt{\frac{4 \pi}{2l+1} \sum_{m=-l}^{l} \left \vert \bar{Q}_{lm} \right \vert^2}~,
\end{eqnarray}
where $\bar{Q}_{lm}$ is defined as
\begin{eqnarray}
\bar{Q}_{lm}=\frac{\sum_{i=1}^{N_c}N_b(i)\bar{q}_{lm}(i)}{\sum_{i=1}^{N_c}N_b(i)}~,
\end{eqnarray}
where $N_c$ is the number of clusters and 
\begin{eqnarray}
\label{stable:eq2}
\bar{q}_{lm}\left(i\right) = \frac{1}{N_b\left(i\right)} \sum_{j=1}^{N_b\left(i\right)} Y_{lm}\left(\hat{\mathbf{r}}_{ij} \right)~.
\end{eqnarray}
Here $N_b(i)$ is the set of bonded neighbors of the $i$-th cluster, the unit vector $ \hat{\mathbf{r}}_{ij}$ specifies the orientation of the bond
between clusters $i$ and $j$, and $Y_{lm}(\hat{\mathbf{r}}_{ij})$ are the corresponding spherical harmonics.

A particularly useful probe of the possible crystal structure of the system is a value of $Q_6$ close to unity 
(see Appendix A of Ref.\ \cite{tenWolde96}). Results for $Q_6$ from the PSW model are
reported in Table \ref{tab:ne1} for the two values of penetrability considered in Fig.\ \ref{fig:fig5}
($\epsilon_a/\epsilon_r=1/4$ and $\epsilon_a/\epsilon_r=1/7$).
{In order to compute $Q_6$,} the center of mass of {each}
 cluster {(as topologically defined before)} is identified. Then,
the cutoff distance for the nearest-neighbors ``bonds'' is selected to be approximately equal
to the second minimum of $g(r)$ ($r\approx 1.5\sigma$). As detailed in Table \ref{tab:ne1}, we find
$0.03 \le Q_6\le 0.1$ for   $\epsilon_a/\epsilon_r=1/4$ ({top} table) and $0.05 \le Q_6\le 0.12$ for $\epsilon_a/\epsilon_r=1/7$ ({bottom} table), depending on the considered values of temperature and density.
These values have been computed with $N=512$ particles but an increase up to $N=1024$ yields only a slight increase {of $Q_6$}.
\begin{table}
\begin{center}
\begin{tabular}{ccccc}
\multicolumn{5}{c}{$\epsilon_a/\epsilon_r=1/4$}\\
\hline
$\rho\sigma^3$ & $k_BT/\epsilon_a$ & $N_c$ & $Q_6$ & $u/\epsilon_a$\\
\hline
0.1 & 1.0 & 13  & 0.04 & -60\\
0.2 & 1.5 & 24  & 0.10 & -57\\
0.3 & 1.7 & 115 & 0.03 & -21\\
0.4 & 1.9 & 132 & 0.05 & -19\\
0.5 & 2.1 & 116 & 0.05 & -18\\
0.6 & 2.4 & 98  & 0.07 & -19\\
0.7 & 2.6 & 84  & 0.04 & -18\\
0.8 & 2.9 & 98  & 0.11 & -19\\
0.9 & 3.2 & 74  & 0.09 & -22\\
1.0 & 3.6 & 67  & 0.05 & -23\\
\hline
\end{tabular}

\vspace{0.5cm}

\begin{tabular}{ccccc}
\multicolumn{5}{c}{$\epsilon_a/\epsilon_r=1/7$}\\
\hline
$\rho\sigma^3$ & $k_BT/\epsilon_a$ & $N_c$ & $Q_6$ & $u/\epsilon_a$\\
\hline
0.1 & 1.0 & 51 & 0.12 & -25\\
0.2 & 1.0 & 39 & 0.06 & -37\\
0.3 & 1.0 & 41 & 0.05 & -37\\
0.4 & 1.0 & 42 & 0.07 & -33\\
0.5 & 1.1 & 50 & 0.29 & -24\\
0.6 & 1.0 & 38 & 0.07 & -36\\
0.7 & 1.7 & 55 & 0.05 & -22\\
0.8 & 2.1 & 58 & 0.11 & -22\\
0.9 & 2.4 & 60 & 0.06 & -21\\
1.0 & 2.8 & 62 & 0.06 & -21\\
\hline
\end{tabular}
\caption{Number of clusters, $Q_6$ parameter, and internal
energy per particle for the non-extensive phases found in the case
$\Delta/\sigma=0.5$ and $\epsilon_a/\epsilon_r=1/4$ ({top} table) and
$\epsilon_a/\epsilon_r=1/7$ ({bottom} table), just below the curves
of Fig.\ \protect\ref{fig:fig5}. The parameter $Q_6$ was calculated on the
final equilibrated particle configuration only, with a
neighbor distance of $1.5\sigma$ in all cases.}
\label{tab:ne1}
\end{center}
\end{table}
Besides $Q_6$, in Table \ref{tab:ne1} we report other properties of the
{blob} phases found with $\Delta/\sigma=0.5$ and
$\epsilon_a/\epsilon_r=1/4$ and $\epsilon_a/\epsilon_r=1/7$, such as the number of clusters
and the internal energy per particle $u/\epsilon_a$.
{We observe that the number of clusters is rather constant (typically $40$--$60$) for penetrability $\epsilon_a/\epsilon_r=1/7$. For the higher penetrability $\epsilon_a/\epsilon_r=1/4$ the number of clusters is generally larger, as expected, but is quite sensitive to the specific density and temperature values. As for the internal energy per particle, we observe that its magnitude is always more than four times larger than the kinetic contribution $\frac{3}{2}k_BT$.}

No conclusive pattern appears from the analysis of results of Table \ref{tab:ne1}, as  there seems to be no well-defined behavior in any of the 
probes as functions of temperature and density, and this irregular
behavior can be also checked by an explicit observation of the corresponding snapshots. 
Nonetheless, these results give no indications of the formation of any regular structure.

The final conclusion of the analysis of the fluid-fluid phase diagram region of the PSW model
is that the system is strictly thermodynamically stable for $\epsilon_a/\epsilon_r < 1/f_{\Delta}$
and strictly {thermodynamically} unstable above it, as dictated by Ruelle's stability criterion.
However, if $\epsilon_a/\epsilon_r > 1/f_{\Delta}$ there exists an intermediate region
where the system looks stable for finite $N$ and becomes increasingly unstable upon approaching
the thermodynamic limit.

The next question we would like to address is whether this scenario persists in the fluid-solid transition,
where already the PS model displays novel and interesting features. This is discussed in the next section.

\section{The fluid-solid transition}
\label{sec:solid}
It is instructive to contrast the expected phase diagram for the SW model
with that of the PSW model.

Consider the SW system with a width $\Delta/\sigma=0.5$ that is a well-studied
intermediate case where both a fluid-fluid and a fluid-solid transition
have been observed \cite{Liu05}. The corresponding schematic phase diagram is displayed
in Fig.\ \ref{fig:fig8} (top panel), where the
critical point is $(k_BT_c/\epsilon_a=1.23, \rho_c\sigma^3=0.309)$ in the temperature-density plane,
and its triple point is
($k_BT_t/\epsilon_a=0.508, P_t\sigma^3/\epsilon_a=0.00003)$ in the temperature-pressure plane, with
$\rho_l\sigma^3=0.835$ and $\rho_s\sigma^3=1.28$ \cite{Liu05}. In Ref.\ \cite{Liu05} no
solid stable phase was found  for temperatures above the triple point, meaning that the
melting curve in the pressure-temperature phase diagram is nearly vertical
(see Fig.\ \ref{fig:fig8}, {top panel}).
\begin{figure}
\begin{center}
\includegraphics[height=12cm]{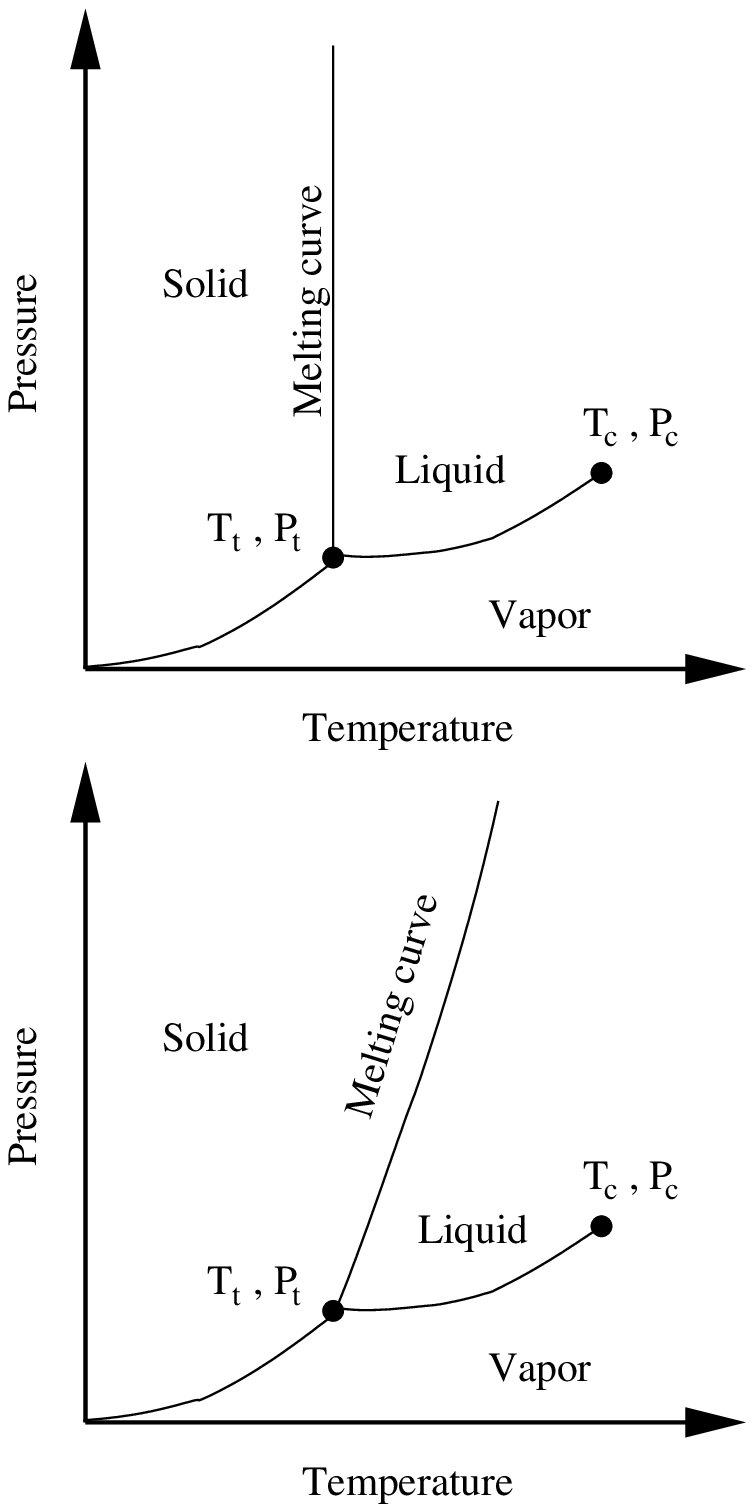}
\end{center}
\caption{Schematic phase diagram of the SW fluid for
$\Delta/\sigma=0.5$ (top panel) and phase diagram of the PSW fluid
for $\Delta/\sigma=0.5$ and $\epsilon_a/\epsilon_r=1/8$ (bottom panel).}
\label{fig:fig8}
\end{figure}
Motivated by previous findings in the fluid-fluid phase diagram,
we consider the PSW model with $\Delta/\sigma=0.5$ and two different penetrability values
$\epsilon_a/\epsilon_r=1/15$ and $\epsilon_a/\epsilon_r=1/8$ in the intermediate region
$1/f_{\Delta} \leq \epsilon_a/\epsilon_r \leq (\epsilon_a/\epsilon_r)_{\text{th}}$ (see Fig.\ \ref{fig:fig3}),
where one expects a normal behavior for finite $N$, but {with} different details depending on the chosen
penetrability. In the present case, the first chosen value ($\epsilon_a/\epsilon_r=1/15$)
lies very close the boundary  ($\epsilon_a/\epsilon_r=1/f_{\Delta}$) of the stability region predicted by Ruelle's criterion, whereas the second
chosen value lies, quite on the contrary, close to the threshold curve $(\epsilon_a/\epsilon_r)_{\text{th}}$.

We have studied the system by isothermal-isobaric (NPT) MC
simulations, with a typical run consisting of $10^8$ MC steps (particle
or volume moves) with an equilibration time of $10^7$ steps. We considered $N=108$ particles and
adjusted the particle moves to have acceptance ratios of {approximately} $0.5$
and volume changes to have acceptance ratios of {approximately} $0.1$.
Note that the typical relaxation time in the
solid region is an order of magnitude higher than that of the
liquid region.

Consider the case $\epsilon_a/\epsilon_r=1/8$ first. The result for the isotherm $k_BT/\epsilon_a=1$ is reported in
Fig.\ \ref{fig:fig9}, this temperature being smaller than the
critical one $k_BT_c/\epsilon_a=1.241$. From this figure we can clearly
see the  jumps in the density corresponding to the gas-liquid coexistence
region and to the liquid-solid coexistence region.
\begin{figure}[ht]
\begin{center}
\includegraphics[height=10cm]{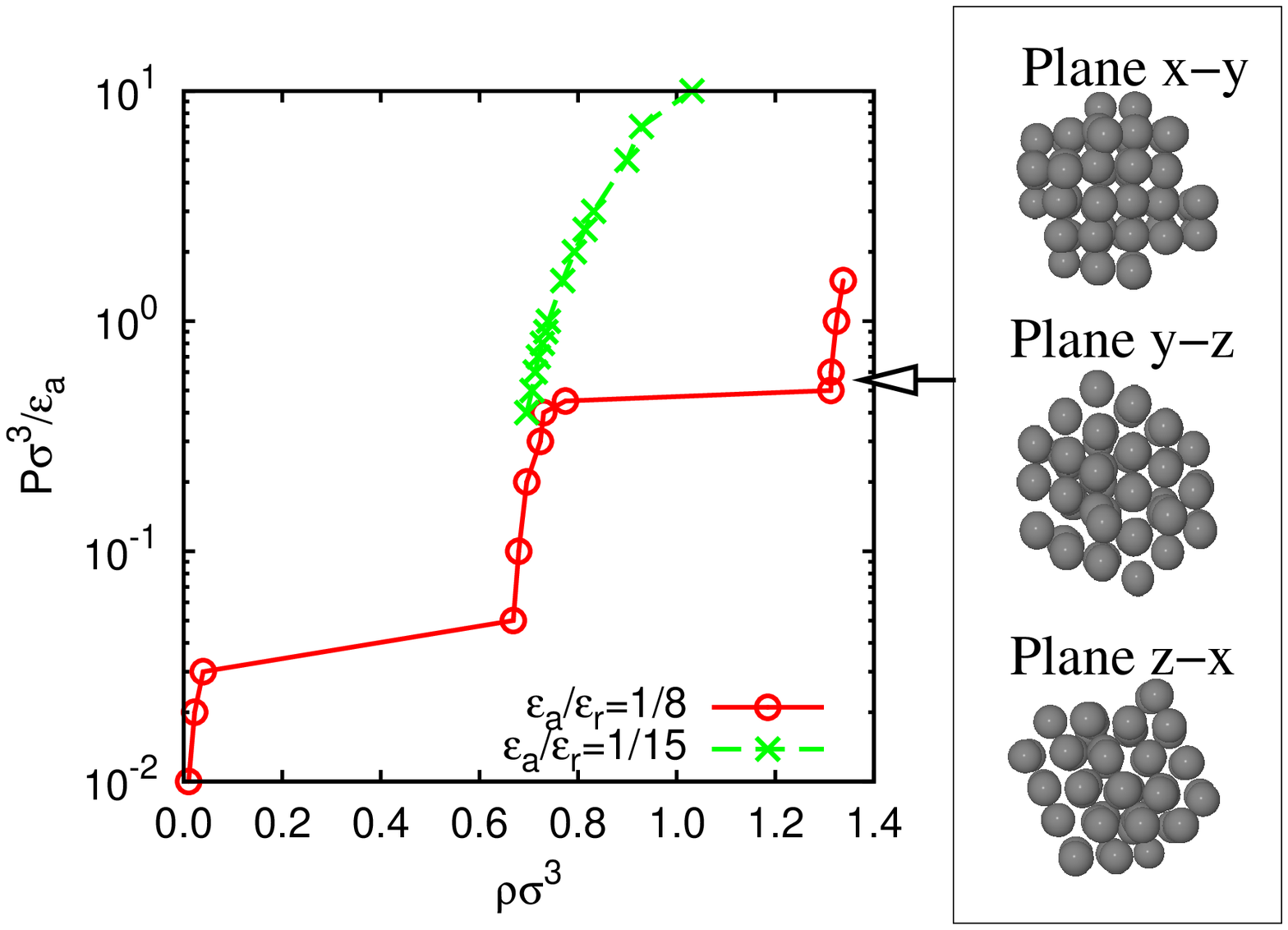}
\end{center}
\caption{Isotherm $k_BT/\epsilon_a=1$ for the PSW system with
$\Delta/\sigma=0.5$ and $\epsilon_a/\epsilon_r=1/8$ and
$\epsilon_a/\epsilon_r=1/15$, as obtained from NPT
MC simulations with $N=108$ particles. The pressure axis is in
logarithmic scale. {Three views of the same snapshot} of the centers of
mass of the clusters in the solid are shown on the right-hand side.}
\label{fig:fig9}
\end{figure}
On the basis of the obtained results, we can foresee a phase diagram
of the PSW system for this particular value of penetrability to be the one sketched
in Fig.\ \ref{fig:fig8} (bottom panel). In particular, the melting curve has a positive
slope in the pressure-temperature phase diagram, {unlike the almost vertical slope of the SW counterpart, as discussed. This implies that penetrability allows for a ``softening'' of the liquid-solid transition, so the liquid and the solid can coexist at a temperature higher than the triple one without the need of a huge increase of pressure.}

Next we also consider a fluid with $\epsilon_a/\epsilon_r=1/15$, just outside the Ruelle
stability region, at the same temperature as before. The results are also reported in Fig.\ \ref{fig:fig8}
and show no indications of a stable solid in the considered range of pressures, in agreement with
the fact that at this very low value of
penetrability the behavior of the system is very close to the SW counterpart.

A specific interesting peculiarity of the PSW system in the intermediate
region $1/f_{\Delta} \leq \epsilon_a/\epsilon_r \leq (\epsilon_a/\epsilon_r)_{\text{th}}$ of Fig.\ \ref{fig:fig3}
is a lack of {full  consistency with known thermodynamic relations}. In this case, in fact, unlike the SW counterpart,
we were unable to trace the coexistence curve between the liquid and the solid using
Kofke's method \cite{Kofke93a,Kofke93b}, which is equivalent to
the numerical integration of the Clausius--Clapeyron equation
\begin{eqnarray}
\label{gibbs-duhem}
\left(\frac{d\ln P}{d\beta}\right)_c=-\frac{\Delta h}{\beta
P\Delta v},\quad \beta\equiv \frac{1}{k_BT},
\end{eqnarray}
with $\Delta h=h_l - h_s$ and $\Delta v=v_l - v_s$, where $h_i$
and $v_i$ denote, respectively, the molar enthalpy and volume of
phase $i$ ($i=l$ for the liquid phase and $i=s$ for the
solid phase); the subscript $c$ indicates that the derivative is
taken along the coexistence line.\
Once a single point on the coexistence curve between the two phases is
known one can use a trapezoid integration scheme \cite{Kofke93b} to
integrate Eq. (\ref{gibbs-duhem}).

In {our} calculation, we have selected {a penetrability $\epsilon_a/\epsilon_r=1/8$ and the isotherm of Fig.\ \ref{fig:fig8}, $k_BT/\epsilon_a=1$, as a reference point. The coexistence pressure at that temperature is $P\sigma^3/\epsilon_a\approx 0.475$ and the molar volume jump is $\Delta v/\sigma^3\approx 1/0.775-
1/1.313\approx 0.529$.} We have then calculated the molar enthalpy in the NPT
ensemble by computing $\langle PV+U\rangle/N$ (where $U$ is the
total {internal energy} of the system) with the result $\Delta h/\epsilon_a\approx
-5.042-(-7.593) = 2.551$. Choosing a spacing in $\beta$ of
$-0.05/\epsilon_a$ we get {from Eq.\ \eqref{gibbs-duhem} a predicted coexistence pressure $P\sigma^3/\epsilon_a\approx 0.789$ at $k_BT/\epsilon_a=1/0.95\simeq 1.053$.} Instead, {however, at the latter temperature} we found the coexistence pressure between $0.5$
and $0.6$. {Despite this quantitative discrepancy, Eq.\ \eqref{gibbs-duhem} is useful to understand that the relatively mild slope of the PSW liquid-solid coexistence line in the pressure-temperature phase diagram is essentially due to the fact that the internal energies of the coexisting liquid and solid phases are not too disparate.}

A close inspection of several snapshots of the obtained solid phase suggests
that, in the intermediate penetrability case, the obtained crystal is
made of clusters of overlapping particles located at the sites of a
regular crystal lattice with $Q_6\approx 0.35$ \cite{tenWolde96}
and a triclinic structure {characterized by a unit cell of sides $a=b=c=\sigma$ and angles
$\alpha=\beta=\pi/3$ and $\gamma=\cos^{-1}(1/4)$ (see three views of a common snapshot in Fig.\ \ref{fig:fig9}).

It is worth stressing that the additional degree of penetrability,
not present in the SW counterpart, is responsible for the coexistence of the
liquid and the solid at not excessively large pressures.
Clearly, we cannot rule out the possibility of other additional solid-solid coexistence regions at
higher pressures.

\section{Conclusions}
\label{sec:conclusions}
In this paper, we have studied the phase diagram of the three-dimensional PSW model. This model combines penetrability, a feature typical of effective potential in complex fluids, with a square-well attractive tail,
accounting for typical effective attractive interactions that are ubiquitous in soft matter. It can then be regarded as the simplest possible model smoothly interpolating between PS {($\epsilon_a/\epsilon_r\to 0$,
$k_BT/\epsilon_r=\text{finite}$)} and SW {($\epsilon_a/\epsilon_r\to 0$, $k_BT/\epsilon_a=\text{finite}$)} fluids, as {one changes penetrability $\epsilon_a/\epsilon_r$ and temperature}.

We have proved that the model is thermodynamically stable when $\epsilon_a/\epsilon_r<1/f_\Delta$, as it satisfies Ruelle's stability criterion \cite{Ruelle69}. Above this value, the fluid is, strictly speaking,
unstable in the thermodynamic limit, exhibiting {non-extensive properties}. For finite $N$, however, it displays a rather rich and interesting phenomenology. In particular, there exists an intermediate region
$1/f_{\Delta} \leq \epsilon_a/\epsilon_r \leq (\epsilon_a/\epsilon_r)_{\text{th}}$ in the {penetrability}-width plane (see Fig.\ \ref{fig:fig3}) where the fluid displays normal or anomalous behavior depending on the
considered temperatures and densities. For sufficiently large temperatures ($T> T_{\text{ins}}(\rho)$) the fluid presents a metastable normal behavior with {(apparently)} stable liquid-liquid and liquid-solid transitions,
provided the relative critical temperatures are above the instability line $T= T_{\text{ins}}$. In this case, we have studied the effect of penetrability on the fluid-fluid transition (see Fig.\ \ref{fig:fig2}) close to the
threshold line $(\epsilon_a/\epsilon_r)_{\text{th}}$ and found that in general the transition {has a} higher critical temperature {than} the SW counterpart. We have attributed this result to the additional degree of
freedom given by penetrability that tends to oppose the formation of a crystal until a sufficient large density is achieved.

Below the instability line {$T_{\text{ins}}(\rho)$}, however, different particles
tend to overlap into a few isolated clusters (blobs) {confined in a small portion of the available volume} and the total energy does no longer scale linearly with
the number of particles $N$. {As a consequence, the fluid becomes thermodynamically unstable} and its properties very anomalous (Fig.\ \ref{fig:fig5}).
The metastable region shrinks as either $\epsilon_a/\epsilon_r$ or $N$ increase (Fig.\ \ref{fig:fig6}).

Above the threshold line $(\epsilon_a/\epsilon_r)_{\text{th}}$ (see Fig.\ \ref{fig:fig3}) {the fluid-fluid coexistence disappears}, since in this case $T_{\text{ins}}$ is too high to allow any phase-separation
(for a given $N$).

An additional interesting feature of the metastable/unstable dualism is included in the hysteresis dependence on the initial condition. When the initial configuration is an unstable one (i.e., a blob) the system melts back to
a normal phase at temperatures that are in generally significantly {higher than} those where the transition normal-to-blob is achieved upon cooling. We have attributed this behavior to the small statistical
weight of the blob configuration in the Boltzmann sampling, in spite of its significantly larger energetic contribution.

We have also studied the fluid-solid transition in the intermediate metastable region
$1/f_{\Delta} \leq \epsilon_a/\epsilon_r \leq (\epsilon_a/\epsilon_r)_{\text{th}}$. We find
that the solid density typically increases with respect to the corresponding SW
case, due to the formation of clusters of overlapping
particles in the crystal sites, as expected on physical grounds. The melting curve is found to
have a {relatively smooth} positive slope, unlike the SW counterpart, and this anomalous behavior is also
reflected in the thermodynamic inconsistency present in the Clausius--Clapeyron thermodynamic equation, thus confirming the
metastable character of the phase. When  penetrability is sufficiently low to be close to the
Ruelle stable region, the system behaves as the corresponding SW system.

One might rightfully wonder whether the finite $N$ metastable phase presented here should have any experimental consequence at all. We believe the answer to be positive. Imagine to be able to craft, through a clever chemical
synthesis process, a fluid that may be described by an effective interaction of the PSW form. Our work has then set the boundary for observing a very intriguing normal-to-collapsed phase by either tuning the temperature/density parameters,
 or by increasing the number of particles in the fluid. In this case, it is the finite $N$ state, rather than the true thermodynamic limit $N \to \infty$, the relevant one.
\section*{Acknowledgements}

R.F. would like to thank Giorgio Pastore for useful discussions on the problem. We thank Tatyana Zykova-Timan and Bianca M. Mladek for enlightening 
discussions and useful suggestions. The support of PRIN-COFIN
2007B58EAB {(A.G.)}, FIS2010-16587 {(A.S)}, and GAAS IAA400720710 {(A.M.)} is acknowledged. Monte Carlo simulations where carried out at the Center for High Performance Computing (CHPC), CSIR Campus, 15 Lower Hope
St., Rosebank, Cape Town, South Africa.


\appendix
\section{Ruelle's stability criterion {in  $d=2$}}
\label{app:Ruelle}
Let us consider the two-dimensional PSW model characterized by
$\epsilon_a/\epsilon_r$ and $\Delta/\sigma<\sqrt{3}-1$.
{The latter condition implies that in a hexagonal close-packed configuration a particle can interact attractively only with its nearest neighbors, so that $f_\Delta=6$.}

Given the number of particles $N$, we want to get the configuration
with the minimum potential energy ${\Phi}_N$. We assume that such a
configuration belongs to the class of configurations described by $m$
rows, each row made of $M$ clusters, each cluster made of $s$
perfectly overlapped particles. The centers of two adjacent clusters
(in the same row or in adjacent rows) are separated a distance
$\sigma$. The total number of particles is $N=mMs$. Figure \ref{fig:fig10}
shows a sketch of a configuration with $m=4$ rows and $M=6$ clusters
per row.
\begin{figure}[tbp]
\begin{center}
\includegraphics[width=8cm]{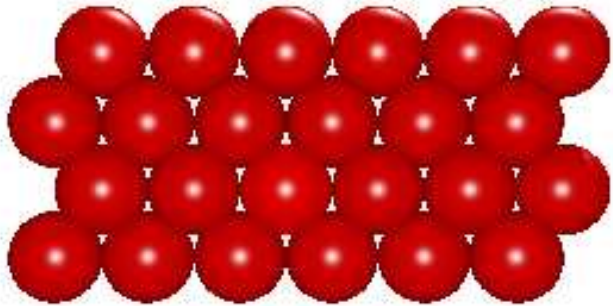}
\caption{Sketch of a configuration with $m=4$ rows and $M=6$ clusters
  per row.\label{fig:fig10}}
  \end{center}
\end{figure}
The potential energy of an individual row is the same as that of the
one-dimensional case \cite{Santos08}, namely
\begin{eqnarray}
{\Phi}^\text{row}=Ms\frac{s-1}{2}\epsilon_r-\left(M-1\right)s^2\epsilon_a.
\label{app:eq1}
\end{eqnarray}
{The first term accounts for the repulsive energy between all possible pairs of particles in a given $s$-cluster,
while the second term accounts for attractions that are  limited to nearest neighbors if $\Delta/\sigma < \sqrt{3}-1$
in $d=2$.}
The potential energy of the whole system is $m{\Phi}^\text{row}$ plus the
attractive energy of nearest-neighbor clusters sitting on adjacent
rows {(and taking into account the special case of boundary rows). The result is}
\begin{eqnarray}
{\Phi}_N(m,s)&=&m\left[Ms\frac{s-1}{2}\epsilon_r-
  \left(M-1\right)s^2\epsilon_a\right]-(m-1)
\left[1+2(M-1)\right]s^2\epsilon_a\nn
&=&N\frac{s-1}{2}\epsilon_r-\left[\frac{3m-2}{m}N-(2m-1)s\right]s\epsilon_a.
\label{app:eq2}
\end{eqnarray}
For a given number of rows $m$, the value of $s$ that minimizes ${\Phi}_N(m,s)$ is found to be
\begin{eqnarray}
s_*(m)&=&N\frac{3m-2}{2m(2m-1)}\left[1-\frac{m}{2(3m-2)}
  \frac{\epsilon_r}{\epsilon_a}\right],
\label{app:eq3}
\end{eqnarray}
which is meaningful only if {$\epsilon_a/\epsilon_r>m/2(3m-2)>1/6$. Otherwise, $s_*(m)=1$. Therefore, the }
corresponding minimum value is
\begin{eqnarray}
{\Phi}_N^*(m)&\equiv&{\Phi}_N(m,s_*(m))\nn
&=&-\frac{N}{2}\epsilon_r\begin{cases}1+N\frac{(3m-2)^2}{2m^2(2m-1)}
\frac{\epsilon_a}{\epsilon_r}\left[1-\frac{m}{2(3m-2)}\frac{\epsilon_r}{\epsilon_a}\right]^2,&\frac{\epsilon_a}{\epsilon_r}>\frac{m}{2(3m-2)},\\
2\left(\frac{3m-2}{m}-\frac{2m-1}{N}\right)\frac{\epsilon_a}{\epsilon_r},&\frac{\epsilon_a}{\epsilon_r}<\frac{m}{2(3m-2)}.
\end{cases}
  \label{app:eq4}
\end{eqnarray}

{Let us first suppose that $\epsilon_a/\epsilon_r<1/6$. In that case, $\epsilon_a/\epsilon_r<m/2(3m-2)$ regardless of the value of $m\geq 1$ and, according to Eq.\ \eqref{app:eq4}, the minimization of ${\Phi}_N^*(m)$  is achieved with $m=M=N^{1/2}$. As a consequence, Ruelle's stability criterion  \eqref{ruelle:eq0} is satisfied in the thermodynamic limit with $B=3\epsilon_a$.}

{Let us now minimize ${\Phi}_N^*(m)$ with respect to $m$ if $\epsilon_a/\epsilon_r>m/2(3m-2)$.} This yields the
quadratic equation $(6-\epsilon_r/\epsilon_a)m^2-12m+4=0$, whose
solution is
\begin{eqnarray}
m_{**}=\frac{2}{3-\sqrt{3+\epsilon_r/\epsilon_a}}.
\label{app:eq5}
\end{eqnarray}
{The condition $\epsilon_a/\epsilon_r>m_{**}/2(3m_{**}-2)$ is easily seen to be equivalent to the condition $\epsilon_a/\epsilon_r>1/6$.}
Therefore, the absolute minimum of the potential energy in that case is
\begin{eqnarray}
{\Phi}_N^{**}&\equiv&{\Phi}_N^*(m_{**})\nn
&=&-\frac{N}{2}\epsilon_r\left[1+\frac{N}{8}\frac{\epsilon_a}{\epsilon_r}\left(3-\sqrt{3+\epsilon_r/\epsilon_a}\right)^3\left(1+\sqrt{3+\epsilon_r/\epsilon_a}\right)\right].
\label{app:eq6}
\end{eqnarray}
The corresponding value of $s_*$ is
\begin{eqnarray}
s_{**}&\equiv&s_*(m_{**})\nn
&=&\frac{N}{4}\left(3-\sqrt{3+\epsilon_r/\epsilon_a}\right)^2.
\label{app:eq7}
\end{eqnarray}
Comparison between Eqs.\ \eqref{app:eq5} and \eqref{app:eq7} shows that
$N=m_{**}^2s_{**}$, i.e., the number of clusters per row equals the
number of rows, $M_{**}=m_{**}$, as might have anticipated by symmetry
arguments.

{Equation \eqref{app:eq6} shows that, if $\epsilon_a/\epsilon_r>1/6$, $\lim_{N\to\infty} (-{\Phi}_N^{**})/N=\infty$ and thus Ruelle's stability condition \eqref{ruelle:eq0} is not fulfilled.}

We could have restricted to a symmetric arrangement  from the very
beginning, i.e., $m=M$ and $N=M^2s$, in which case Eq.\ \eqref{app:eq2} yields
\begin{eqnarray}
{\Phi}_N(M,s=N/M^2)&=&M^2s\frac{s-1}{2}\epsilon_r-
\left(3M^2-4M+1\right)s^2\epsilon_a\nn
&=&\frac{N}{2}\left(\frac{N}{M^2}-1\right)\epsilon_r-
\left(3M^2-4{M}+1\right)\frac{N^2}{M^4}\epsilon_a.
\label{app:eq8}
\end{eqnarray}
The minimum value {(if $\epsilon_a/\epsilon_r>1/6$)} corresponds to the value $M=m_{**}$ given by
Eq.\ \eqref{app:eq5}, as expected.


\end{document}